\documentclass[aps,prl,reprint,twocolumn,superscriptaddress,floatfix,longbibliography]{revtex4-1}
\usepackage{graphicx,amsmath,amsfonts,amssymb,amsthm,xr}
\usepackage{epsfig,amsmath,amssymb,color,dsfont,upgreek,physics}
\usepackage{mathrsfs}
\usepackage{mathtools}
\usepackage{bbold}
\usepackage{comment}
\usepackage{soul}
\usepackage{float}

\usepackage[bookmarks=true,colorlinks,linkcolor=OrangeRed,urlcolor=NavyBlue,citecolor=RoyalBlue]{hyperref}
\usepackage[dvipsnames]{xcolor}
\usepackage{orcidlink}

\definecolor{mygold}{rgb}{0.93,0.69,0.13}

\definecolor{mypurple}{rgb}{0.49,0.18,0.56}

\definecolor{mygreen}{rgb}{0,0.5,0}

\definecolor{mygreen}{rgb}{0,0.5,0}
\definecolor{myred}{rgb}{0.7,0,0}
\definecolor{myblue}{rgb}{0,0,0.5}

\begin{document}
\title{Disorder-Free Localization as a Purely Classical Effect}
\author{Pablo Sala${}^{\orcidlink{0000-0001-7512-505X}}$}
\email{psala@caltech.edu}
\affiliation{Department of Physics and Institute for Quantum Information and Matter,
California Institute of Technology, Pasadena, California 91125, USA}
\affiliation{Walter Burke Institute for Theoretical Physics, California Institute of Technology, Pasadena, California 91125, USA}
\affiliation{Department of Physics, Technische Universit\"at M\"unchen, James-Franck-Straße 1, D-85748 Garching, Germany}
\author{Giuliano Giudici}
\email{giuliano.giudici@physik.uni-muenchen.de}
\affiliation{Department of Physics and Arnold Sommerfeld Center for Theoretical Physics (ASC), Ludwig-Maximilians-Universit\"at M\"unchen, Theresienstra\ss e 37, D-80333 M\"unchen, Germany}
\affiliation{Munich Center for Quantum Science and Technology (MCQST), Schellingstra\ss e 4, D-80799 M\"unchen, Germany}
\author{Jad C.~Halimeh${}^{\orcidlink{0000-0002-0659-7990}}$}
\email{jad.halimeh@physik.lmu.de}
\affiliation{Department of Physics and Arnold Sommerfeld Center for Theoretical Physics (ASC), Ludwig-Maximilians-Universit\"at M\"unchen, Theresienstra\ss e 37, D-80333 M\"unchen, Germany}
\affiliation{Munich Center for Quantum Science and Technology (MCQST), Schellingstra\ss e 4, D-80799 M\"unchen, Germany}

\begin{abstract}
Disorder-free localization (DFL) is an ergodicity breaking mechanism that has been shown to occur in lattice gauge theories in the quench dynamics of initial states spanning an extensive number of gauge superselection sectors. Whether DFL is intrinsically a quantum interference effect or can arise classically has hitherto remained an open question whose resolution is pertinent to further understanding the far-from-equilibrium dynamics of gauge theories. In this work, we utilize cellular automaton circuits to model the quench dynamics of large-scale quantum link model (QLM) formulations of $(1+1)$D quantum electrodynamics, showing excellent agreement with the exact quantum case for small system sizes. Our results demonstrate that DFL persists in the thermodynamic limit as a purely classical effect arising from the finite-size regularization of the gauge-field operator in the QLM formulation, and that quantum interference, though not a necessary condition, may be employed to enhance DFL.
\end{abstract}

\date{\today}
\maketitle

\textbf{\textit{Introduction.---}}The pursuit of a general theoretical framework of the far-from-equilibrium dynamics of quantum many-body systems is a major goal in condensed matter physics \cite{Eisert2015,Rigol_review,Deutsch_review,Buca2023unified}.
Whereas generic interacting many-
body models are expected to thermalize according to the eigenstate thermalization hypothesis (ETH) \cite{Deutsch1991,Srednicki1994}, it has become clear that several ETH violations exist. Examples include integrable systems \cite{Rigol_2008}, systems with quantum many-body scars \cite{Moudgalya2018,Turner2018}, Hilbert space fragmentation~\cite{Sala_PRX,khemani_localization_2020}, and many-body localization (MBL) \cite{Basko2006,Gornyi2005,Nandkishore_review,Abanin_review,Alet_review}. The latter was first predicted to exist in disordered systems. However, it is now known that the presence of disorder is not a necessary ingredient, and without it localization can still arise. For example, so-called Stark MBL involves adding a strong tilted potential in a clean system of interacting fermions \cite{vanNieuwenburg2019,Schulz2019}, which has also been experimentally demonstrated to lead to a strong suppression of dynamics in cold-atom and trapped-ion quantum simulators \cite{Scherg2020,Morong2021}. Another mechanism for generating MBL without disorder appears in quantum many-body models with local symmetries, known as gauge theories, which are fundamental frameworks of modern physics that describe the interactions of elementary particles as mediated by gauge bosons \cite{Weinberg_book,Gattringer_book}. The principal property of gauge theories are their \textit{local gauge invariance}, which encodes the laws of nature through intrinsic relations between the local distribution of matter and the surrounding electric fields, as exemplified through Gauss's law in quantum electrodynamics (QED). Upon preparing the system in a superposition over an extensive number of the gauge superselection sectors and subsequently performing a global quench, localized dynamics can emerge where the system retains memory of its initial state \cite{Smith2017,Brenes2018}. This can occur even when the quench Hamiltonian is nonintegrable, disorder-free, and translation-invariant with a homogeneous initial state. This ergodicity-breaking mechanism, known as \textit{disorder-free localization} (DFL), has been demonstrated in various models \cite{smith2017absence,Metavitsiadis2017,Smith2018,Park2019,Russomanno2020,Papaefstathiou2020,McClarty2020,hart2021logarithmic,Zhu2021,karpov2021disorder,Sous2021,Halimeh2021enhancing,Chakraborty2022,Gao2023}, and has generally been attributed to an emergent effective disorder in the Hilbert space associated with the background charges corresponding to the spanned gauge superselection sectors.

In Brenes \textit{et al.}~\cite{Brenes2018}, DFL was studied in the Schwinger model, where Gauss's law was employed in order to integrate out the gauge fields, resulting in a purely fermionic model with long-range Coulomb interactions and an explicit correlated-disorder term related to the background charges. The gauge-coupling constant then controls the disorder and interaction strengths. Soon thereafter, it was shown that DFL also persists, at least for small system sizes, at zero gauge coupling for \textit{quantum link model} (QLM) regularizations of the Schwinger model where the gauge and electric-field operators are represented by spin-$S$ operators \cite{Halimeh2021stabilizingDFL}. Interestingly, it was also recently shown that starting in thermal ensembles spanning an extensive number of gauge superselection sectors, quench dynamics can also give rise DFL in these QLMs and also in $\mathbb{Z}_2$ gauge theories \cite{Halimeh2022TDFL}. These works have thus suggested that DFL can occur without an explicit disorder term arising in some exact mapping or perhaps without the need of quantum interference effects. This has hence raised again the question as to the origins of DFL. Given that the quantum simulation of gauge theories has recently become a very active research area in quantum many-body physics, and experimental realizations abound \cite{Bernien2017,Kokail2019,Martinez2016,Muschik2017,Klco2018,Schweizer2019,Goerg2019,Mil2020,Klco2020,Yang2020,Zhou2022,Nguyen2021,Wang2021,Mildenberger2022,Wang2022}, this motivates a better theoretical understanding of exotic nonergodic gauge-theory dynamics that can be probed on such platforms.

In this work, we ask whether DFL in lattice gauge theories can arise \textit{purely classically}, at least in some form, or whether quantum interference effects are a necessary ingredient for it to occur. We approach this question by modeling the dynamics of spin-$S$ $\mathrm{U}(1)$ QLMs using cellular automaton circuits, which have been employed in several recent works on systems with unconventional symmetries~\cite{Iaconis_2019,Feldmeier_anomal, morningstar2020_kinetic,Iaconis_2021,Hart_2022,2022arXiv221116788P,https://doi.org/10.48550/arxiv.2210.15607,2022PhRvB.106i4303F,Sala_21}. Here we show that DFL persists in the thermodynamic limit as a purely classical effect. We attribute this form of DFL to the regularized \emph{finite} structure of gauge superselection sectors in QLMs, leading to reducible dynamics~\cite{ Ritort03,Garrahan_review,Sala_PRX,khemani_localization_2020,Moudgalya_review}. Furthermore, we show excellent quantitative agreement in the imbalance and infinite-temperature correlations with the quantum case through exact diagonalization (ED) on small system sizes. We argue that this type of DFL is distinct from, though can occur concomitantly with, the type connected to a finite gauge coupling in the Schwinger model.

\begin{figure}[t!]
 \includegraphics[scale=0.5]{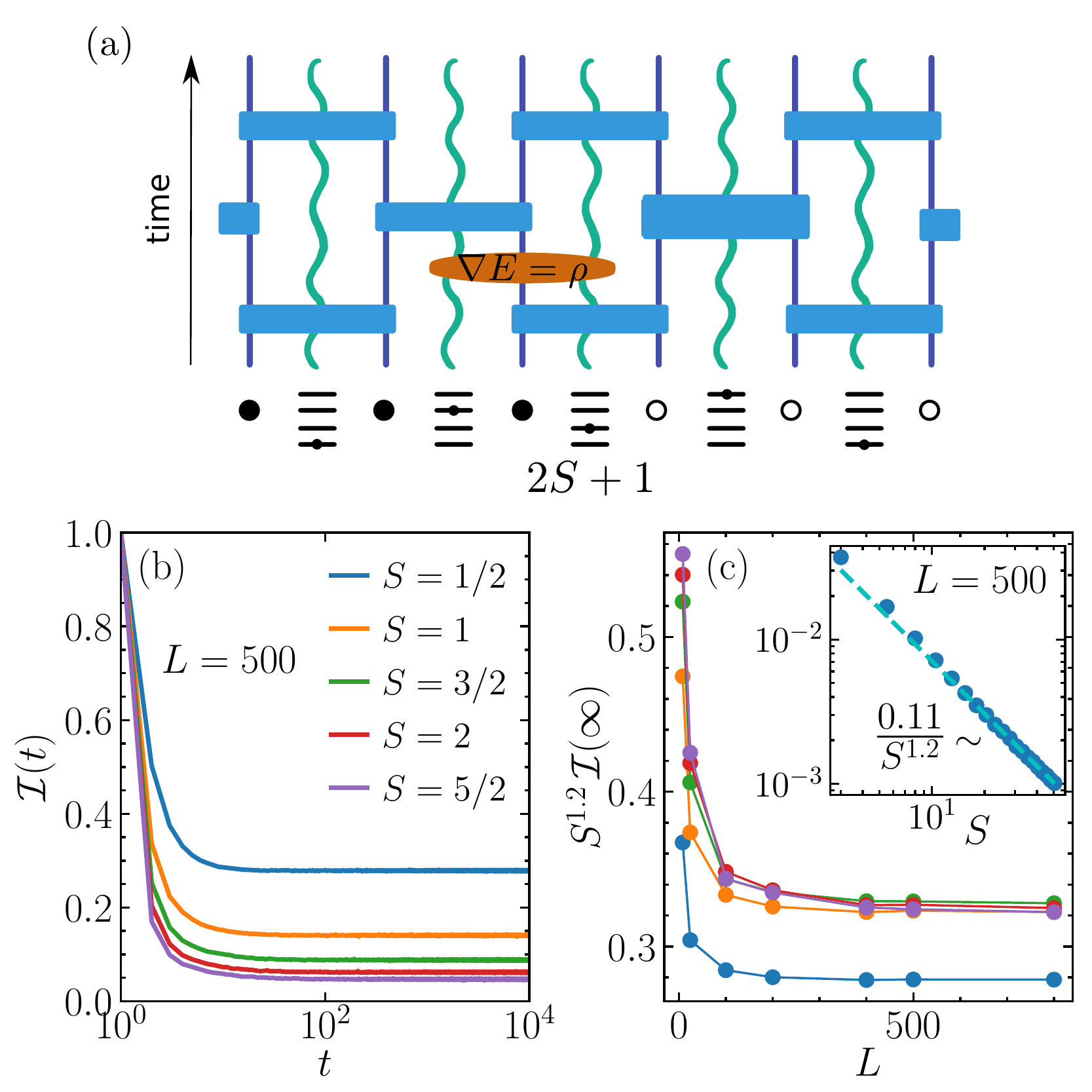}
\caption{(Color online). \textbf{Disorder-free localization as a purely classical effect.} (a) A schematic of the cellular automaton implementation of the spin-$S$ $\mathrm{U}(1)$ quantum link model~\eqref{eq:QLM} and the enforcement of Gauss's law~\eqref{eq:Gj}. Purple straight lines indicate matter sites and their discrete time evolution, and green wiggly lines denote the links on which electric and gauge fields reside. The matter occupation on a site is either $0$ or $1$, while the electric field takes on a value in $\{-S,\ldots,S\}$. (b) The dynamics of the imbalance~\eqref{eq:imbalance} after starting in a domain-wall state of $L=500$ sites and averaging over $N= 8000$ randomly chosen electric-field configurations, modeling a superposition of an extensive number of gauge superselection sectors in the quantum case. Disorder-free localization arises for all considered values of $S$. Throughout the paper, we have used $N=10000$ for system sizes $L<200$ and $N=8000$ for $L\geq200$. (c) Infinite-time value of the imbalance shows convergence with system size for $L\gtrsim400$, indicating persistence in the thermodynamic limit with a power-law behavior in $S$, as shown in the inset.}
 \label{fig1}
\end{figure}

\textbf{\textit{Model and diagnostics.---}}A numerically and experimentally relevant regularization of QED on a lattice takes the form of a \textit{quantum link} formulation of its gauge and electric-field operators, whereby they are represented by spin-$S$ operators \cite{Chandrasekharan1997,Wiese_review,Kasper2017}. Lattice QED is then approached in the Kogut--Susskind limit of $S\to\infty$, although it is known that the low-energy physics of lattice QED is faithfully reproduced already for small values of $S$ on quantum devices \cite{Zache2021achieving,Halimeh2022achieving}. The corresponding model is the spin-$S$ $\mathrm{U}(1)$ quantum link model (QLM) with Hamiltonian \cite{Chandrasekharan1997,Wiese_review,Kasper2017,Hauke2013}
\begin{align}\nonumber
\hat{H}=\sum_{j=1}^L\bigg[&\frac{J}{2\sqrt{S(S+1)}}\big(\hat{\sigma}^-_j\hat{s}^+_{j,j+1}\hat{\sigma}^-_{j+1}+\text{H.c.}\big)\\\label{eq:QLM}
&+\frac{\mu}{2}\hat{\sigma}^z_j+\frac{\kappa^2}{2}\big(\hat{s}^z_{j,j+1}\big)^2\bigg],
\end{align}
where $L$ is the number of sites. Throughout this work, periodic boundary conditions (PBC) are employed and the lattice spacing is set to unity. The matter fields are represented by the Pauli matrices $\hat{\sigma}^z_j$ on sites $j$, while the gauge and electric fields are respectively represented by the spin-$S$ operators $\hat{s}^+_{j,j+1}/\sqrt{S(S+1)}$ and $\hat{s}^z_{j,j+1}$ on the links between adjacent sites $j$ and $j+1$. The coupling constant $J=1$ sets the overall energy scale, $\mu$ is the fermionic mass, and $\kappa$ is the gauge-coupling strength. The spin-$1/2$ formulation has been experimentally realized in large-scale implementations using Rydberg atoms \cite{Bernien2017,Surace2020} and a tilted Bose--Hubbard superlattice \cite{Yang2020,Zhou2021,Wang2022}. The spin-$S$ $\mathrm{U}(1)$ QLM hosts a $\mathrm{U}(1)$ gauge symmetry with generator
\begin{align}\label{eq:Gj}
   \hat{G}_j=(-1)^j\bigg(\hat{s}^z_{j-1,j}+\hat{s}^z_{j,j+1}+\frac{\hat{\sigma}^z_j+\mathds{1}}{2}\bigg),
\end{align}
with eigenvalues $g_j$, so-called background charges, where $(-1)^jg_j$ are consecutive integers in $\{-2S,\ldots,2S{+}1\}$. A gauge superselection sector in the total Hilbert space is defined as a unique set of these eigenvalues $\mathbf{g}=\{g_1,g_2,\ldots,g_L\}$. It is worth noting here that not all charge configurations are physical for QLM regularizations, and, as we will see later, this truncation of the local Hilbert space of the gauge and electric fields will have consequences on DFL that get more pronounced with smaller $S$.

In the following, we will investigate DFL in this model by computing the matter imbalance $\mathcal{I}(t)=\sum_{j=1}^L \langle \hat{\sigma}^z_j(0) \rangle \langle \hat{\sigma}^z_j (t) \rangle/L$,
starting from a domain-wall initial state where the matter occupies only one half of the system while the gauge degrees of freedom are at infinite temperature. We will also compute the two-point unequal-time correlation function $\langle \hat{\sigma}^z_j(t) \hat{\sigma}^z_0(0) \rangle$, where both matter and gauge degrees of freedom are initially at infinite temperature. In particular, we will employ discrete-time (blocked) cellular automaton circuits (CAC) to model the dynamics of these quantities in order to ascertain whether DFL can arise in this inherently classical setup. A further advantage of using CAC is that it allows accessing large system sizes and long evolution times.

\textbf{\textit{Cellular automaton circuits.---}}Previous works have utilized CAC to probe the hydrodynamic behavior of (classical) systems with kinetic constraints and different kinds of unconventional conservation laws such as dipole-moment conservation~~\cite{Iaconis_2019,Feldmeier_anomal, morningstar2020_kinetic,Iaconis_2021,Hart_2022,2022arXiv221116788P,https://doi.org/10.48550/arxiv.2210.15607,2022PhRvB.106i4303F,Sala_21}. The method is schematically depicted in Fig.~\ref{fig1}a. Analogously to the quantum Hamiltonian in Eq.~\eqref{eq:QLM}, we distinguish between matter sites (straight purple lines) of index $j$ and gauge links (curly green lines) between sites, lying on a chain of $L$ sites. The (classical) discrete spin $s_{j,j+1}$ representing the discretized local electric field on the link between sites $j$ and $j+1$ takes on values $s_{j,j+1} \in \{-S,\dots,S\}$, while the matter occupation on site $j$ is given by $\sigma_j \in \{-1,+1\}$, denoted in Fig.~\ref{fig1}a as $\circ$ and $\bullet$, respectively. Consequently, the state of the system at any time-step $t$ is given by a string configuration $C(t)=(\sigma_1, s_{1,2}, \sigma_2,\dots,s_{L-1,L}, \sigma_L, s_{L,1})$, explicitly assuming PBC.

The dynamics is governed by local gates $P_{j,j+1}$ acting on pairs of neighboring sites $j$ and $j+1$ and their intermediate link, akin to the $3$-local terms in the Hamiltonian~\eqref{eq:QLM}.
Using an analogous formulation to that in Ref.~\cite{Sala_21}, we define the action of a local gate by two integers $\alpha$ and $\beta$, such that upon applying $P^+_{j,j+1}= \{\alpha,-\beta\}$ or its inverse $P^-_{j,j+1}=\{-\alpha,\beta\}$, the matter and gauge degrees of freedom are updated as $\sigma_{j} \to \sigma_{j} \pm \alpha$, $\sigma_{j+1} \to \sigma_{j+1} \pm \alpha$ and $s_{j,j+1} \to s_{j,j+1} \mp\beta$, respectively. These updates are randomly applied among those for which $|\sigma_{j} \pm \alpha|= 1$, $|\sigma_{j+1} \pm \alpha|= 1$, and $|s_{j,j+1} \pm \beta| \leq S$, with symmetric transition rates such that detailed balance is satisfied for the uniformly random ensemble over string configurations $C$. As a result, a local gate $P_{j,j+1}$ simply implements a random permutation between two allowed strings $C\to P_{j,j+1}C$. A (discrete) time-step is then given by a sequence of such non-overlapping gates acting on neighboring sites and the links in between as shown in Fig.~\ref{fig1}a. The full time evolution follows from the application of several such layers of gates. In the following, we consider the case $\alpha=2\beta=2$, to appropriately model the tunneling term of Hamiltonian~\eqref{eq:QLM}.

Furthermore, we define the ``classical'' Gauss's law as $G_j=(-1)^j\big[s_{j-1,j}+s_{j,j+1}+(\sigma_j+1)/2\big]$ at site $j$ and its two neighboring links; cf.~Eq.~\eqref{eq:Gj}. One can then see that evaluating $G_j$ on a string $C$ as given by $G_j(C)$, agrees with $G_j(P_{i,i+1}C)$ for all $i,j=1,\dots, L$. This is trivially true for $i\neq j-1, j$. In the remainder cases, one finds $G_j(P_{i,i+1}C)=G_j(C)\pm (\beta-\alpha/2)$, which is equal to $G_j(C)$ for $\alpha=2\beta$. Hence, the previously described CAC leaves the values of $G_j$ invariant for all $j$ and at all times, leading to reducible classical dynamics whose sectors correspond to the gauge superselection sectors of the spin-$S$ $\mathrm{U}(1)$ QLM~\eqref{eq:QLM}. This motivates us to investigate whether DFL, well-established for the spin-$S$ $\mathrm{U}(1)$ QLM, can also arise in this \textit{purely classical} setting, in which case we can adjudge whether quantum interference effects are necessary for DFL to arise. 

\begin{figure}[t!]
 \includegraphics[scale=0.5]{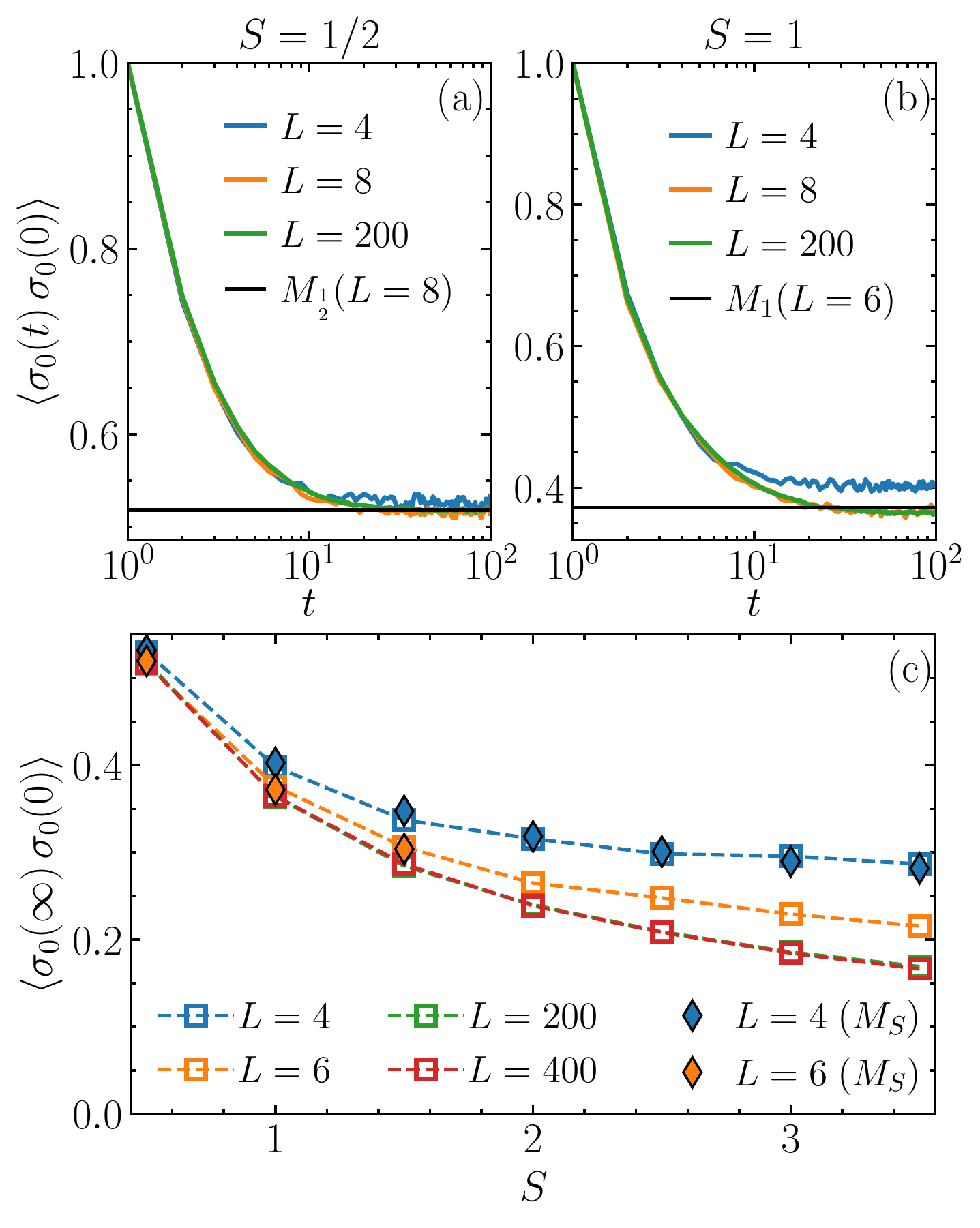}
\caption{(Color online). \textbf{Localized dynamics of the matter autocorrelator.} Modeling of the dynamics of the matter autocorrelator for (a) $S=1/2$ and (b) $S=1$ for using cellular automaton circuits. The black horizontal line demarcates the corresponding Mazur bound~\eqref{eq:MazurBound}. (c) Infinite-time value of the matter autocorrelator, showing a monotonic decrease with $S$, and convergence to the thermodynamic limit for system sizes $L\gtrsim 200$ sites. Diamonds represent the finite-size Mazur's bounds obtained from Eq.~\eqref{eq:MazurBound}. }
 \label{fig2}
\end{figure}

\textbf{\textit{Localized dynamics from CAC.---}}We now focus on the dynamics as calculated through CAC. We prepare an initial string $C(0)$ with a domain-wall structure in the matter degrees of freedom, i.e., $\sigma_j(0)=\pm1$ on the left (right) half of the chain, and a uniform random configuration of the electric field variables $s_{j,j+1}$. We then let it evolve and average the observables of interest $A$ over $N$ such different preparations $\langle A \rangle \equiv \sum_{\{C(t)\}}A(C(t))/N$. The imbalance is then computed as follows
\begin{align}\label{eq:imbalance}
    \mathcal{I}(t)=\frac{1}{L}\sum_{j=1}^L\langle\sigma_j(0)\sigma_j(t) \rangle.
\end{align}
The corresponding dynamics is shown in Fig.~\ref{fig1}b for a system of $L=500$ sites and for $N=8000$ for various values of $S$. We see that the imbalance settles into a plateau of finite value at intermediate times that persists for all investigated timescales ($t\leq 10^4$ time-steps). Hence, we see clear features of DFL. These results can be considered to be in the thermodynamic limit, as shown by the finite-size scaling of the plateau value of $\mathcal{I}(t)$ in Fig.~\ref{fig1}c, which demonstrates convergence with system size for $L\gtrsim400$ at all considered values of $S$. The plateau also exhibits a power-law decay with $S$ at sufficiently large $S$, as shown in the inset of Fig.~\ref{fig1}c for $L=500$ sites, indicating a vanishing value in the ``Kogut-Susskind'' limit $S\to \infty$. We also note that the imbalance dynamics shows very good agreement between CAC and ED for small system sizes upon breaking energy conservation by Trotterizing the dynamics (see Supplemental Material \cite{SM}).

\begin{figure}[t!]
 \includegraphics[scale=0.5]{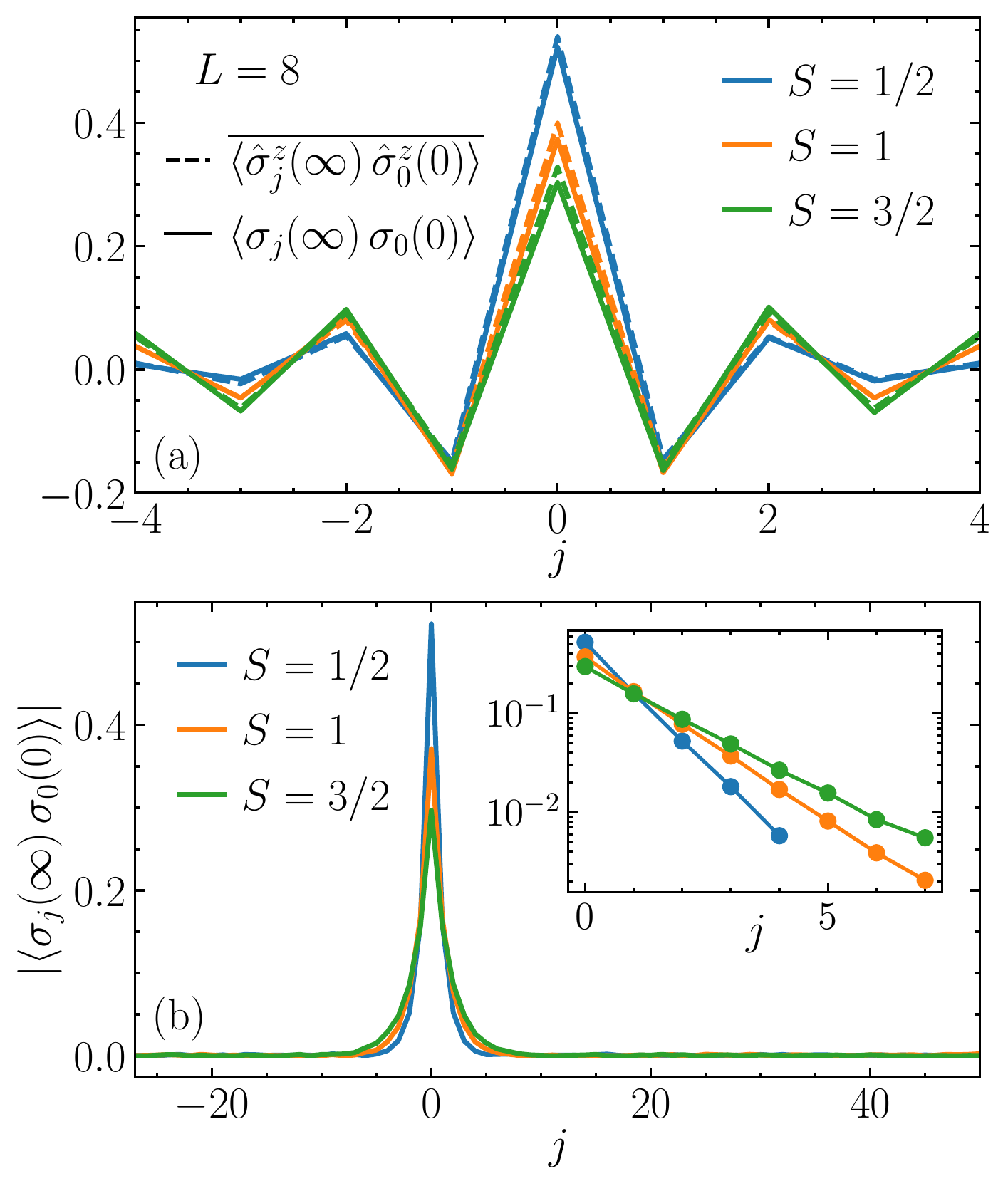}
\caption{(Color online). \textbf{Correlation spatial profile.} (a) Spatial profile of the two-point unequal-time correlator at $t\to\infty$, obtained through CAC (ED) for the classical (quantum) case. For all considered values of $S$, we find excellent agreement between the classical and quantum cases. (b) Spatial profile of the two-point unequal-time correlator calculated from CAC at $t\to\infty$ for $L=200$ sites, showing exponential localization for all considered values of $S$. The inset shows that localization is more prominent with decreasing $S$.}
 \label{fig3}
\end{figure}

We now turn to the computation of the matter two-time correlation function $\langle \sigma_j(t) \sigma_0(0) \rangle$, where we initialize the circuit in a state with both matter and gauge degrees of freedom in a uniform random configuration. The discrete dynamics of the equal-space two-time function for $S=1/2$ and $1$ is plotted for various values of $L$, indicating convergence to the thermodynamic limit as shown in Fig.~\ref{fig2}a-b. Its long-time value is shown in Fig.~\ref{fig2}c for $L$ up to $400$ sites as a function of $S$. The saturation value attained by the correlation function for $t \to \infty$ can be lower-bounded by making use of the conserved quantities via Mazur's bound $M_S$~\cite{Mazur69}. Such a bound holds for both quantum and classical systems as long as one evaluates the correlations on stationary states of the dynamics~\cite{Mazur69,Dhar_2021}, and their numerical values agree when the same set of conserved quantities are considered. In this case, one can show that
\begin{align}\label{eq:MazurBound}
    \lim_{T\to \infty}\frac{1}{T}\int_0^Tdt \langle \hat{\sigma}^z_0(t)\hat{\sigma}_0^z(0) \rangle{\geq}\sum_\mathbf{g} \frac{ \mathrm{Tr} \big\{\hat{\sigma}^z_0 \hat{P}_\mathbf{g}\big\}^2 }{\mathrm{Tr}\big\{\hat{P}_\mathbf{g}\big\}}{\equiv}M_S,
\end{align}
where $\hat{P}_\mathbf{g}$ is the projector onto the gauge sector labeled by the background-charge distribution $\mathbf{g}$. In fact, the same approach was used to show a finite saturation value of infinite-temperature correlations in the presence of strong fragmentation of the Hilbert space~\cite{Sala_PRX}, as well as for boundary correlations~\cite{Rakovszky_slioms,Sala_21,Lehmann_22}. In practice, we compute $M_S$ by scanning over all superselection sectors and taking full advantage of the fact that the projectors $\hat{P}_{\mathbf{g}}$ are diagonal. We observe that its value mildly depends on the system size $L$ and provides a tight bound on equal-space two-time correlations at infinite time, as shown in Fig.~\ref{fig2}.

We now compare the CAC results to those obtained from ED for the quantum model~\eqref{eq:QLM} at $\mu=\kappa=0$. For this purpose, we focus on the infinite-time two-point matter classical and quantum correlators $\langle \sigma_j(t) \sigma_0(0) \rangle$ and $\langle \hat{\sigma}^z_j(t) \hat{\sigma}^z_0(0) \rangle$, respectively, at $t\to\infty$, shown in Fig.~\ref{fig3}a for $L=8$ sites at several values of $S$. For the computation of quantum correlators, we exploit dynamical typicality and obtain infinite temperature averages starting from random initial pure states \cite{dyn_typ}. We observe excellent agreement between the stationary states of the classical and quantum cases. In this spirit, we employ CAC to calculate this spatial profile for $L=200$ sites and several values of $S$ in Fig.~\ref{fig3}b, showing exponential localization in all considered cases. This quantitative agreement corroborates the equivalent propagation of local perturbations under CAC and quantum evolution at infinite temperature, which is halted due to the interplay of an extensive number of superselection sectors in the dynamics.

Given that the CAC computation is inherently classical, and yet we see DFL for all $S$, we are lead to conclude that the occurrence of DFL in spin-$S$ $\mathrm{U}(1)$ QLMs does not require quantum interference. Instead, it is in this case a purely classical effect that we attribute to the regularized finite structure of gauge sectors at finite values of $S$. Indeed, Fig.~\ref{fig1}b-c show that the imbalance plateau takes on a value that decreases with $S$, while Figs.~\ref{fig2}c and \ref{fig3}b show that the peak in the long-time spatial profile of the matter two-time function is also a decreasing function of $S$. These observations indicate that in the limit of $S\to\infty$, DFL originating due to this classical effect is not expected to emerge. 

\begin{figure}[t!]
 \includegraphics[scale=0.5]{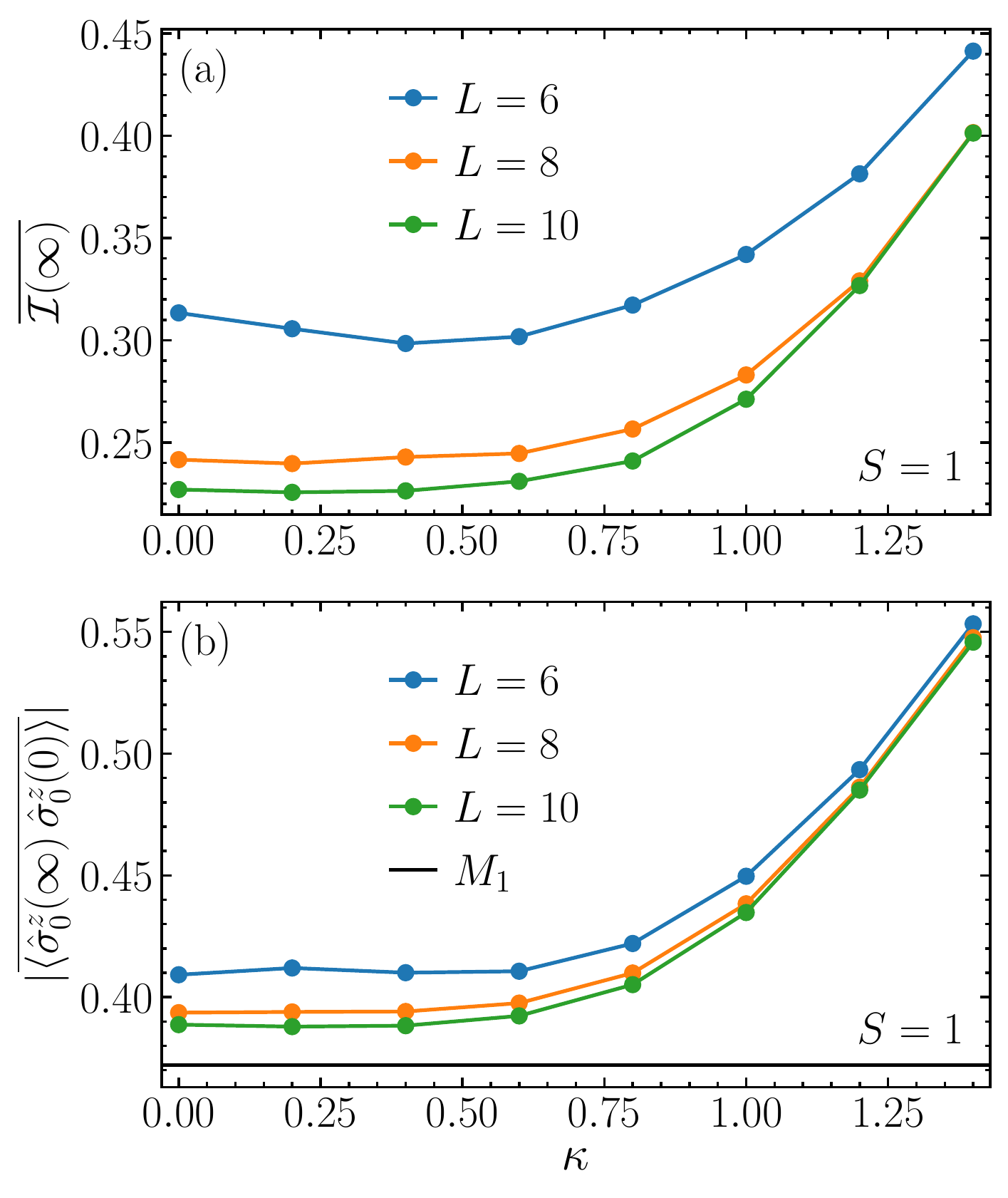}
\caption{(Color online). \textbf{Effect of gauge coupling.} (a) Plateau value of the imbalance as a function of the gauge coupling $\kappa$ for $S=1$ in the quantum case, calculated in ED. At sufficiently large system size $L$, we find a monotonic increase with $\kappa$ in the plateau value. (b) The equal-space two-time correlator at $t\to\infty$ as a function of $\kappa$ as computed in ED for the quantum case. Similarly to the imbalance plateau value, it shows a monotonic increase with $\kappa$ at sufficiently large $L$. Mazur's bound $M_1$ is shown with a straight black horizontal line.}
 \label{fig4}
\end{figure}

We have found that the CAC computation we employ here faithfully models the stationary state of the infinite-temperature correlation-function dynamics generated by the Hamiltonian~\eqref{eq:QLM} for $\mu=\kappa=0$. In the case of the lattice Schwinger model achieved in the limit $S=\infty$, it was argued that a finite $\kappa$ is a necessary condition for DFL to emerge \cite{Brenes2018}. We have shown that, for finite $S$, this is not the case, and the finite regularization of the gauge sectors in the spin-$S$ $\mathrm{U}(1)$ QLM suffices for DFL to emerge. 

We now study the effect of $\kappa$ on DFL in the spin-$S$ $\mathrm{U}(1)$ QLM following the analysis of previous works \cite{Brenes2018,Halimeh2021stabilizingDFL}. For this purpose, we focus on the case of $S=1$ in Fig.~\ref{fig4} \footnote{For $S=1/2$, the gauge-coupling term is a inconsequential energetic constant.}. Using ED, we calculate the quench dynamics of the imbalance starting in an initial state with a domain-wall structure in the matter fields and gauge degrees of freedom at infinite temperature. We find that the plateau value of the imbalance, shown in Fig.~\ref{fig4}a, monotonically increases with $\kappa$ at sufficiently large values of $L$, noting that it is actually finite for $\kappa=0$. Similarly, the equal-space two-time correlator at $t\to\infty$ increases with $\kappa$, as shown in Fig.~\ref{fig4}b. This confirms the general conclusion of Ref.~\cite{Brenes2018} that $\kappa$ acts as a parameter for disorder strength enhancing the localization. However, we note that at finite $S$ their mapping from the lattice Schwinger model to an interacting fermionic system with correlated disorder is no longer exact.

\textbf{\textit{Discussion and outlook.---}}Using cellular automaton circuits, we were able to show that disorder-free localization due to the combination of several superselection sectors can arise in the spin-$S$ $\mathrm{U}(1)$ quantum link model in the thermodynamic limit purely from the finite regularization of its gauge sectors, without the need for finite gauge coupling, which is necessary for DFL to occur in the lattice Schwinger model limit at $S=\infty$. We validated our conclusions through modeling the dynamics of the imbalance and infinite-temperature two-point unequal-time correlation functions, showing excellent agreement with results from the exact diagonalization for the full quantum model at small system sizes.

Since CAC is an inherently classical setup, this shows that quantum interference is not a necessary condition for localization when combining extensively many superselection sectors. This is related to other forms of localization in the absence of disorder, such as in the context of Hilbert-space fragmentation when combining several fragments. Since DFL has been established as an intriguing ergodicity breaking paradigm in gauge theories, a thorough understanding of its origins can shed light into the far-from-equilibrium dynamics of such models.

An interesting avenue for future work would be to understand how prominent classical DFL in higher-dimensional systems is \cite{Osborne2023}. Another avenue involves exploring the role of the emergent random potential in the Schwinger model in competition with the long-range Coulomb potential, and how necessary each of those terms are to observe DFL. It would be also interesting to understand the thermal properties of each of the involved superselection sectors, which combined lead to finite correlations, and investigate the specific constrained structure arising at finite $S$. In the case of dipole-conserving systems, the cause of this behavior has been associated with the presence of statistically localized degrees of freedom labeling all Krylov subspaces left invariant by the dynamics~\cite{Rakovszky_slioms, Commutant_algebras}.

\bigskip

\begin{acknowledgments}
The authors are grateful to Johannes Knolle, Julius Lehmann, Yahui Li, Frank Pollmann, and Tibor Rakovszky for enlightening discussions and for previous collaborations on related topics. This work was supported by the Walter Burke Institute for Theoretical Physics at Caltech, and the Institute for Quantum Information and Matter. J.C.H.~acknowledges funding from the European Research Council (ERC) under the European Union’s Horizon 2020 research and innovation programm (Grant Agreement no 948141) — ERC Starting Grant SimUcQuam, and by the Deutsche Forschungsgemeinschaft (DFG, German Research Foundation) under Germany's Excellence Strategy -- EXC-2111 -- 390814868. G.G.~acknowledges support from the Deutsche Forschungsgemeinschaft (DFG, German Research Foundation) under Germany’s Excellence Strategy – EXC2111 – 390814868 and from the ERC grant QSIMCORR, ERC-2018-COG, No.~771891.
\end{acknowledgments}

\bibliography{biblio}

\onecolumngrid
\newpage

\appendix

\begin{center}
    {\Large Supplemental Material \\ 
    }
\end{center}

 \section{CAC vs non-Hamiltonian quantum dynamics}
\begin{figure}[h!]
\centering
    \includegraphics[scale=0.5]{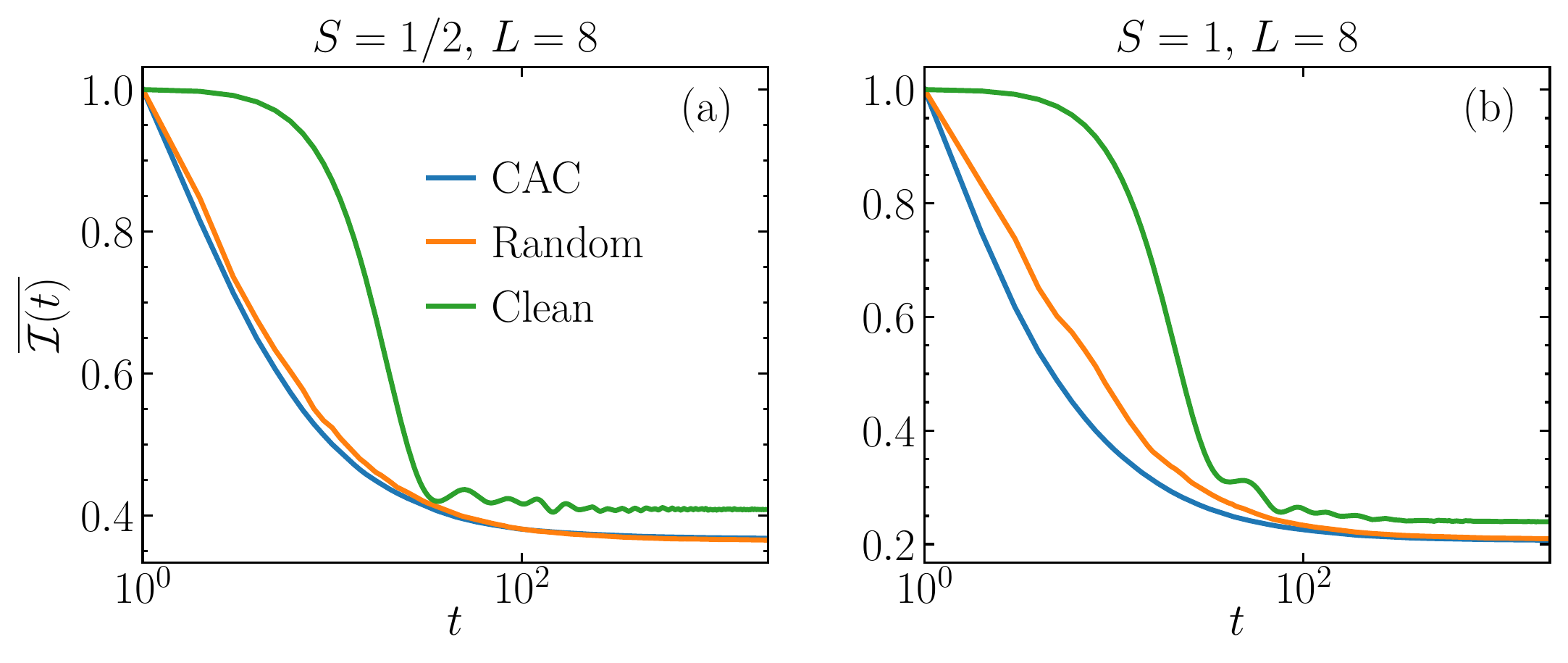}
    \caption{ CAC dynamics (blue) vs. quantum dynamics generated by the QLM Hamiltonian Eq.~\eqref{eq:QLM} (green) and random quantum dynamics obtained from the unitary circuit Eq.~\eqref{eq:RUC} (orange) for $S=1/2$ (a) and $S=1$ (b).  }
    \label{fig1_SM}
\end{figure}

In the main text we showed how CAC is capable of reproducing quantitatively the spatial profile of the infinite-temperature matter two-time function in the long-time limit of the quantum dynamics generated by the Hamiltonian \eqref{eq:QLM}. However, as we discuss now, we did not find the same quantitative agreement for the asymptotic value of the imbalance $\mathcal{I}(t)$. CAC and clean quantum dynamics are depicted in Fig.~\ref{fig1_SM} for $L=8$ and $S=1/2,1$. In the quantum case $\mathcal{I}(t)$ saturates to a non-zero value, which is approximated from below by the CAC dynamics. We attribute this discrepancy to the lack of energy conservation in the CAC dynamics and confirm this hypothesis by comparing the latter to the non-Hamiltonian dynamics generated by the random quantum circuit 
\begin{equation}
\label{eq:RUC}
U_{j,j+1} = e^{-i \eta_{j,j+1} \hat{h}_{j,j+1} } ,
\end{equation}
where $\hat{h}_{j,j+1}$ is the QLM Hamiltonian density on two matter sites plus one link and $\eta_{j,j+1}$ is a Gaussian random variable with zero mean and variance equal to one.
Fig.~\ref{fig1_SM} shows the result obtained by averaging over $10$ realizations of this random unitary dynamics and exhibits perfect quantitative agreement for the long-time value attained by the imbalance.

\end{document}